# A natural language interface to a graph-based bibliographic information retrieval system

Yongjun Zhu[1], Erjia Yan, Il-Yeol Song

College of Computing and Informatics, Drexel University, 3141 Chestnut Street, Philadelphia, PA 19104.

{zhu, ey86, song}@drexel.edu

**Abstract**

With the ever-increasing scientific literature, there is a need on a natural language interface to bibliographic information retrieval systems to retrieve related information effectively. In this paper, we propose a natural language interface, NLI-GIBIR, to a graph-based bibliographic information retrieval system. In designing NLI-GIBIR, we developed a novel framework that can be applicable to graph-based bibliographic information retrieval systems. Our framework integrates algorithms/heuristics for interpreting and analyzing natural language bibliographic queries. NLI-GIBIR allows users to search for a variety of bibliographic data through natural language. A series of text- and linguistic-based techniques are used to analyze and answer natural language queries, including tokenization, named entity recognition, and syntactic analysis. We find that our framework can effectively represents and addresses complex bibliographic information needs. Thus, the contributions of this paper are as follows: First, to our knowledge, it is the first attempt to propose a natural language interface to graph-based bibliographic information retrieval. Second, we propose a novel customized natural language processing framework that integrates a few original algorithms/heuristics for interpreting and analyzing natural language bibliographic queries. Third, we show that the proposed framework and natural language interface provide a practical solution in building real-world natural language interface-based bibliographic information retrieval systems. Our experimental results show that the presented system can correctly answer 39 out of 40 example natural language queries with varying lengths and complexities.

---

[1] Corresponding author

## 1. Introduction

Bibliographic information retrieval systems such as Web of Science, Scopus, and Google Scholar have become an unalienable component in searching bibliographic data (Chadegani et al., 2013). These systems continuously index ever-increasing scientific literature, thus providing a source for scholars to learn, create, and represent new knowledge (Jacso, 2005). These systems, however, have several methodological limitations. Web of Science and Scopus provide form-based interfaces, in which users first select fields (e.g., topic and author) and then type appropriate values for each field. Advanced searching is an option, in which users can formulate queries using tags and Boolean operators (Score, 2009). While form-based interfaces are effective for simple queries (e.g., *papers written by an author*), it is known that the interface is not adequate in dealing with ad-hoc information and imposes a burden on users to understand and select fields or tags (Noessner et al., 2010). Apart from form-based search interfaces, Google Scholar leverages an intuitive keyword-based interface (Falagas et al., 2008). While a keyword-based search interface is perhaps the most widely used for everyday information retrieval, it has challenges in understanding queries, particularly in recognizing how keywords are related to one another (Tumer et al., 2009).

An alternative to these form- and keyword-based interfaces is a natural language interface. A natural language interface allows users to formulate queries expressed in natural language. It is more flexible than a form-based interface and has a higher level of expressiveness than a keyword-based interface (Androutsopoulos et al., 1995). In current form- and keyword-based bibliographic information retrieval systems, users are provided with limited options to represent complex bibliographic queries (e.g.,

*papers on information retrieval, which were cited by John's papers in SIGIR*). Through a natural language interface, users can represent complex bibliographic queries using natural language and get relevant results in one step without the need to fill out forms or try with different keywords. While form- and keyword-based interfaces are predominately adopted in current bibliographic information retrieval systems, we design a framework of developing a natural language interface for bibliographic information retrieval. In particular, we aim to address the following two core questions concerning natural language bibliographic information retrieval:

1) How to design a framework to build a system that can interpret, query, and answer natural language bibliographic queries; and

2) How to implement and evaluate a bibliographic information retrieval system with a natural language interface?

The framework interprets bibliographic queries expressed in controlled natural language and returns relevant bibliographic data and relations. Natural language queries supported in the framework are restricted to complex nominal phrases that describe bibliographic entities. We implement a natural language-based interface on a graph-based bibliographic information retrieval system designed in our previous work (Zhu, Yan, & Song, 2016). While our previous work introduced a general framework for a graph-based bibliographic information retrieval system called GIBIR, the current work focuses on the interface on top of the GIBIR. Theoretically, this study is novel because it introduces natural language interfaces for graph-based bibliographic information retrieval. This study examines a series of approaches including query interpretation, processing, and visualization specifically for bibliographic searching environment—they consider a wide range of bibliographic information needs and the characteristics of bibliographic data. Thus, a natural language interface tailored for bibliographic environment provides a new and effective way of searching bibliographic data. In addition, from practical aspects, by enabling users to formulate bibliographic information needs in natural language, it liberates users from learning cumbersome ways of representing those needs. With ever-increasing bibliographic data, a natural language interface allows an effective retrieval of data by enabling the representation of complex bibliographic information needs and simplifying the search process into a single step without multiple refining procedures.

This paper makes the following three contributions: First, it is the first attempt to propose a natural language interface to graph-based bibliographic information retrieval. While there was an attempt to design a natural language interface (Doszkocs and Rapp, 1979) a few decades ago, it was proposed for retrieving bibliographic data specific to MEDLINE. In this paper, we propose a novel approach utilizing practical natural language techniques for modern graph-based bibliographic information retrieval systems. Second, we propose a novel customized natural language processing framework that integrates a few original algorithms/heuristics for interpreting and analyzing natural language bibliographic queries. The proposed framework has been developed by carefully examining characteristics of bibliographic data and bibliographic queries and thus, provides higher accuracy to satisfy more bibliographic information needs. Third, we show that the proposed framework and natural language interface provide a practical solution to build real-world natural language interface-based bibliographic information retrieval systems. Our experimental results show that the presented system can correctly answer 39 out of 40 example natural language queries with varying lengths and complexities

The remainder of this paper is organized as follows: Section 2 surveys background knowledge on natural language interface, named entity recognition, and syntactic analysis, which form the foundation of the framework. Section 3 presents the natural language processing framework in detail. Section 4 shows the implemented system with example queries and reports experimental results of the framework by testing 40 natural language queries. Section 5 concludes our paper.

## 2. Literature Review

We review three areas of research that directly connect to natural language-enabled bibliographic information retrieval systems, including natural language interface, named entity recognition, and syntactic analysis.

*2.1. Natural language interface*

Natural language interfaces (NLI) are used to query structured information stored in databases. Two types of NLI can be distinguished: one is natural language interfaces to databases (NLIDB), in which a relational database is used to store structured information; the other is natural language interfaces to knowledge bases (NLIKB) that use an ontology to manage information (e.g., Habernal & Konopík, 2013; Abacha & Zweigenbaum, 2015). While the two types of NLI use different database systems, they have common components, including the interpretation of natural language queries and concept mappings between entities in queries and databases (e.g., Cafarella & Etzioni, 2005; Tablan et al., 2008).

The relational data model (Codd, 1970) proposed in the early 1970s had a major impact on NLIDB research. NLIDB are highly portable and can be attached to existing databases because relational databases are the norm of most traditional information retrieval systems (Vicknair et al., 2010). Compared to NLIDB, NLIKB have a relatively short history with the inception of semantic web (Berners-Lee et al., 2001). Databases in this category deploy rich expressive power of ontologies represented in the resource description framework (Miller, 1998), thus generally achieving higher performances (e.g., Kaufmann & Bernstein, 2010). Readers can refer to Androutsopoulos and colleagues' work (1995) for a comprehensive review of NLIDB systems. Recent NLIKB systems include PowerAqua (Fazzinga & Lukasiewicz, 2010), ORAKEL (Cimiano et al., 2008), FREyA (Damljanovic et al., 2010), PANTO (Wang et al., 2007), and NLP-Reduce (Kaufmann et al., 2007).

The NLI designed in this paper is an NLI to graph databases (e.g., Roy & Zeng, 2013). Graph databases have comparable expressive power with ontologies (i.e., triple stores), but a much higher scalability, which are more suitable to real-world systems (Angles & Gutierrez, 2008). Graph databases have been increasingly used in information retrieval systems (e.g., Park & Lim, 2015). Graph databases excel relational databases in answerable questions due to its advantage on representing complex relations among data given that natural language queries are represented using complex relations among concepts. Given the graph-like characteristics of bibliographic data as discussed in our previous work (Zhu, Yan, & Song, 2016), a natural language interface to graph database-based bibliographic information retrieval systems provides a novel way of accessing and retrieving bibliographic data.

*2.2. Named entity recognition*

Named entity recognition (NER) is a task of identifying names of things in texts. These things include but not limited to persons, organizations, locations, and biomedical entities (Nadeau & Sekine, 2007). Early NER systems used rule-based methods to recognize named entities. In a rule-based NER system, patterns in a text are identified and appropriate rules are handcrafted based on those patterns. Thus, a rule-based method is mainly used in self-contained domains and has a limited applicability (e.g., Rau, 1991). A dictionary-based NER system utilizes predefined dictionaries and performs a look-up in texts (e.g., Ryu, Jang, & Kim, 2014; Mu, Lu, & Ryu, 2014). The method is widely used in domains such as biomedicine, in which named entities are well recorded and managed, for instance, in protein recognition (Tsuruoka & Tsujii, 2003) and drug recognition (Rindflesch et al., 2000). Another popular category of NER is statistical NER (e.g., Derczynski et al., 2015). Widely used statistical NER includes maximum entropy (ME)- (Chieu & Ng, 2002), hidden Markov models (HMM)- (Bikel et al., 1997), and conditional random fields (CRF)-based (McCallum & Li, 2003) NER systems. Some NER systems use more than one type of NER: for example, Stanford NER (Finkel et al., 2005) provides both dictionary- and statistical-based NER through a gazette feature.

Bibliographic data are relatively easy to obtain through well-known bibliographic databases such as Web of Science and DBLP. Thus, in this paper, we used a dictionary-based approach to recognize bibliographic named entities (i.e., authors, papers, organizations, terms, and sources) from a natural language query. By recognizing bibliographic named entities in a query, we are able to extract these entities as well as their relations to learn and answer queries.

*2.3. Syntactic analysis (Parsing)*

A classic way of parsing is to derive parses from a string of words based on a structure grammar of prewritten phrases (i.e., context-free grammar) (e.g., Earley, 1980). With the introduction of annotated data such as The Peen Treebank (Marcus et al., 1993), a number of statistical parsers were proposed and became popular. Readers can refer to Collins' work (1997) for a more extensive review on statistical parsing models.

Two popular ways of representing syntactic structures are constituency and dependency. For constituency, words in a sentence are organized into nested constituents; while for dependency, dependent relations between words are shown (Klein & Manning, 2004). Dependency parses can be obtained from dependency parsers (e.g., Fersini et al., 2014) or phrase structure parsers (i.e., constituency) by a conversion system (e.g., De Marneffe et al., 2006). The proposed framework uses a dependency structure to identify grammatical relations among words. Because we are interested in grammatical relations among bibliographic named entities recognized in natural language queries, dependency structures are more straightforward than constituency structures that also show relations between phrases.

## 3. The Framework

The framework is designed to take a natural language query as the input and return correct answers as the output. This is achieved by translating the input into a database query language. A natural language query is translated into a graph query language because we use a graph database to manage bibliographic data. Multiple steps are involved in the translation, including finding answers to questions such as: 1) what is being asked? 2) what entities should be used to constrain the answer? and 3) how does the asked entity relate to other entities? Figure 1 uses a flow chart to describe how the core components of the framework interact with each other.

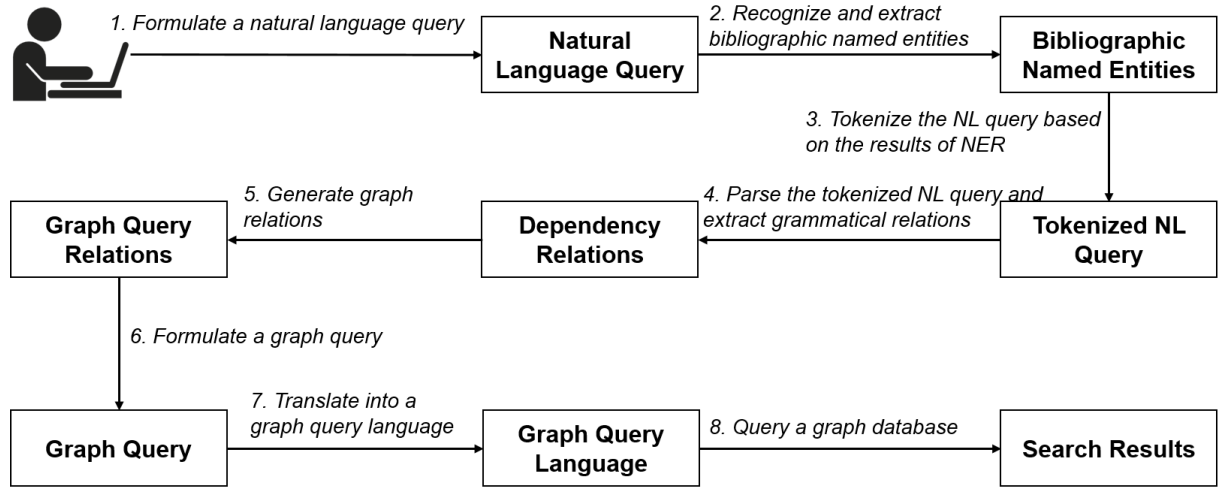

Figure 1. The flow chart of the designed framework

The steps are as follows: 1) a user formulates a query expressed in natural language; 2) bibliographic named entity recognition is performed by referencing predefined dictionaries and recognized bibliographic named entities are then extracted; 3) a natural language query is tokenized based on the result of bibliographic named entity recognition; 4) the tokenized natural language query is parsed to identify grammatical relations among bibliographic named entities; 5) the grammatical relations are filtered and graph relations are generated ; 6) a graph query is formulated by combining bibliographic named entities and graph relations; 7) the graph query is translated into a graph query language; and 8) a graph database is queried.  In the following subsections, we introduce each step in detail with example queries.

3.1. *The formulation of natural language queries*

Although the framework is designed to process a natural language query, it is not a question answering system. Thus, a complete sentence with an interrogative pronoun is not supported in the framework. Instead, noun phrases such as "*papers that were written by John*" and "*authors of papers that were published in SIGIR*" are expected queries. Because the interpretation of natural language queries depends on syntactic analysis, queries are expected to have no grammatical error. In addition, relative pronouns, such as "that", are expected to be included in a query to guarantee that a syntactic parser parses the query correctly. For example, a query "*papers that were written by John*" is the preferred form of "*papers written by John*".

*3.2. The recognition and extraction of bibliographic named entities*

We adopt a dictionary-based named entity recognition approach. We use a simple map structure to construct a dictionary, in which keys are names of bibliographic entities (e.g., "*John*", "*SIGIR*", and "*information retrieval*") and values are their bibliographic types (i.e., *Paper*, *Author*, *Term*, *Source*, and *Organization*). These five bibliographic types are regarded as the most useful as shown in previous studies (e.g., Sun, Yu, and Han 2009). A dictionary is constructed by preprocessing the bibliographic dataset on which we perform searches. Five types of bibliographic instances and their type information are extracted from a self-explanatory dataset. Disambiguation is not performed due to the lack of appropriate identification data. We also add five bibliographic types as keys with annotations to show that they are bibliographic types. For example, the entry <*"paper", "class_Paper"*> is added to the dictionary so that the system recognizes words such as "*paper*" and "*author*" in natural language queries. An additional annotation "*class_*" is added because we want to differentiate five entity types with bibliographic entities.

An approximate string matching algorithm introduced in Gusfield's work (1997) is used to implement the NER algorithm. In the algorithm, a distance of 1 was assigned to insertion, deletion, and substitution of a character. A maximum distance of 1 was allowed, so that we can recognize plurals or singulars when we have only one form of the two of bibliographic named entities. For example, "*Information System*" in a query could be identified as a named entity when we only have the term "*Information Systems*" in our dictionary

*3.3. The tokenization of natural language queries*

We tokenize queries based on the results of named entity recognition to prepare parsing in the next step. After recognizing named entities, we mark named entities of multiples words as single tokens, and then feed queries into a standard tokenizer. This supervised tokenization complements tokenizers' shortage of domain knowledge on technical terms. For example, without using the results of named entity recognition, terms composed of multiple words such as "*information retrieval*" will be processed into two different tokens. Tokenization based on the results of named entity recognition can avoid this limitation because terms recognized as a single named entity are treated as one token. Table 1 shows the difference between tokenization without NER and with NER using an example query "*papers about information retrieval and data mining*", in which tokens are separated by pairs of parentheses.

Table 1. Tokenization without NER and with NER

| Query | *papers about information retrieval and data mining* |
|---|---|
| **Tokenization without NER** | *(papers), (about), (information), (retrieval), (and), (data), (mining)* |
| **Tokenization with NER** | *(papers), (about), (information retrieval), (and), (data mining)* |

*3.4. The parsing of tokenized natural language queries and the extraction of grammatical relations*

We use Stanford parser (Klein & Manning, 2003) to parse queries. The output we generate is the Stanford dependencies (De Marneffe et al., 2006) that use 56 grammatical relations to represent binary relations among tokens. Grammatical relations are used to find out which tokens depend on or modify other tokens. For a bibliographic natural language query, parsing is used to find out grammatical relations among bibliographic named entities represented by tokens. Table 2 shows the dependency relations of a sample query "*papers about information retrieval and data mining*". Readers can refer to De Marneffe and colleague's work (2006) for a detailed explanation of each dependency relation.

Table 2. Dependency relations of the query "papers about information retrieval and data mining"

| Order | Subject | Object | Relation Code | Relation Name |
|---|---|---|---|---|
| 1 | | papers | root | root |
| 2 | information retrieval | about | case | case marker |
| 3 | papers | information retrieval | nmod | nmod_preposition |
| 4 | information retrieval | and | cc | coordination |
| 5 | papers | data mining | nmod | nmod_preposition |
| 6 | information retrieval | data mining | conj | conj_collapsed |

For queries that involve citations such as "*papers about information retrieval that were cited by papers that were written by John*", they are divided into two parts: a cited part and a citing part. By doing so, we reduce the complexities and errors in interpreting queries, because a long list of dependency relations may be error-prone. By dividing the example query into two parts, we no longer need to consider grammatical relations between "*papers*" in the cited part and "*John*" in the citing part. This is a practical way to improve the performance of a parser, and thus, words such as "cited", "cites", "cite", and "citing" are used to divide a query into two parts. Parsing is separately applied to each part, and the results are integrated in a later step to generate graph relations.

*3.5. The generation of graph relations from dependency relations*

A graph query is a graph representation of a natural language query, in which nodes are recognized bibliographic named entities and links are relations of those entities. Graph relations denote relations that are necessary for building complete graph queries that represent natural language queries. Thus, graph relations are subsets of dependency relations, and graph relations are selected from dependency relations. Irrelevant relations (i.e., relations among non-bibliographic named entities) that are included in dependency relations are omitted in this process. The selection is performed by considering both the patterns of queries and the database schema that is used to store bibliographic data.

Figure 2 shows the algorithm we use to select graph relations from dependency relations. We build the heuristics by combing the test results of a list of expected queries and the database schema. Thus, the heuristics introduced here are dependent on the database schema we use and subject to change if a different schema is employed (see Figure 1 in Zhu, Yan, and Song, 2016) for the schema we used in the graph-based system).

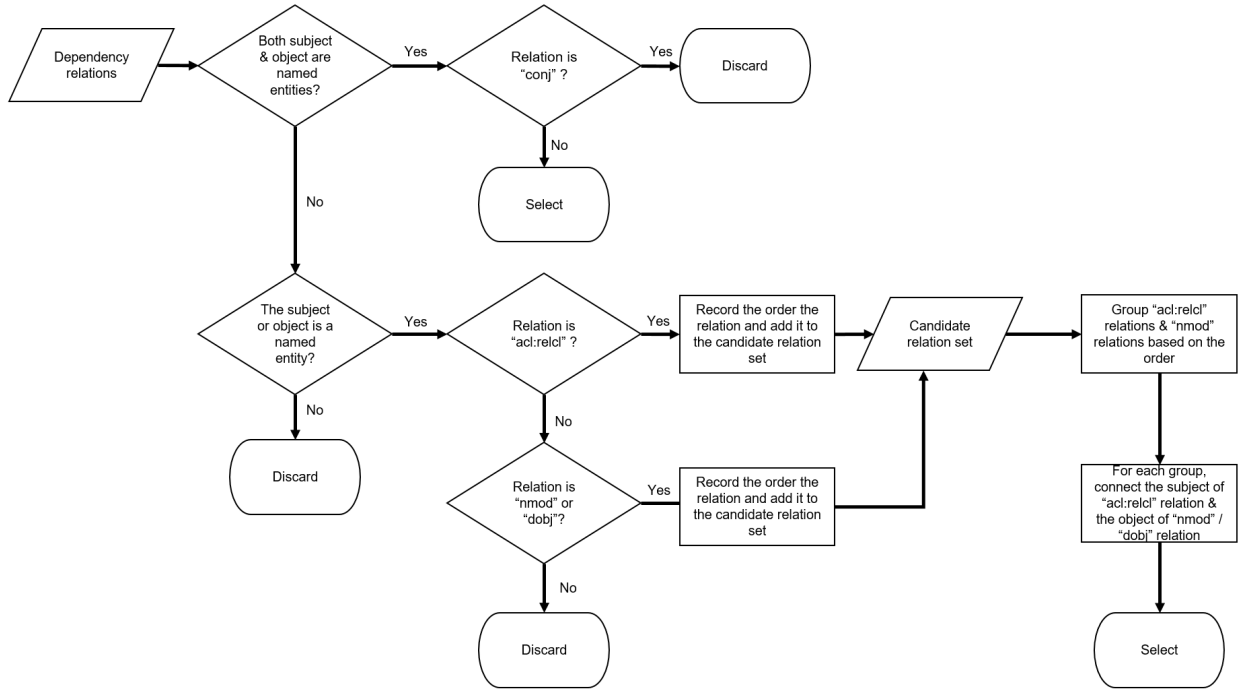

Figure 2. The flow chart of selecting graph relations from dependency relations

As shown in Figure 2, a relation is selected as a graph relation if both the subject and the object of the relation are named entities. "conj" denotes "conjunct", and it is used if two tokens are connected by a coordinating conjunction, such as "and" and "or". In our case, the relations do not play constructive role in building a graph query, and is thus discarded. Accordingly, the third and fifth relations in Table 2 are selected as graph relations while the sixth relation is not. Table 3 shows another dependency relations of an example query "*papers that were written by John*".

Table 3. Dependency relations of the query "papers that were written by John"

| Order | Subject | Object | Relation Code | Relation Name |
|---|---|---|---|---|
| 1 | | papers | root | root |
| 2 | written | papers | nsubjpass | nominal passive subject |
| 3 | papers | that | ref | referent |
| 4 | written | were | auxpass | passive auxiliary |
| 5 | papers | written | acl:relcl | relative clause modifier |
| 6 | John | by | case | case marker |
| 7 | written | John | nmod | nmod_preposition |

Table 3 shows the case in which two bibliographic entities are not directly connected by a dependency relation. It is a normal use case and the algorithm can deal with such use cases. First, the fifth and seventh dependency relations are selected. Then, the subject of fifth relation "*papers*" and the object of seventh relation "*John*" are connected to form a new graph relation as shown in Figure 2. It is a repeated pattern in bibliographic natural language queries that two relation types "acl"relcl" and "nmod" are used to connect two bibliographic named entities.

*3.6. The formulation of graph queries*

In this step, bibliographic named entities are converted into graph nodes, and graph relations are checked for connectedness and direction. We also integrate a cited part and a citing part for queries that involve citations in this step.

3.6.1. The conversion of bibliographic named entities to graph nodes

The conversion takes place in three steps. First, we identify the bibliographic named entity that a query is asking. For example, in the query "*papers that were written by John*", the answer node is "*papers*". The identification of an answer node is to locate the object of a "root" relation in parsing results (e.g., "*papers*" in Table 3). Second, we assign each bibliographic named entity a unique instance name that will be used when generating a graph query language. This allows us to differentiate bibliographic named entities with the same name and type the entity "*papers*" in the query "*papers that were cited by papers that were written by John*". Lastly, we identify bibliographic named entities that constrain the answer node. For example, "*information retrieval*" in the query "*papers about information retrieval*" constrains the answer node "*papers*" by adding a condition. If the type of a bibliographic named entity does not contain the string "*class_*", the named entity is a constraint node. This explains the reason that we add the string "*class_*" to the values of five bibliographic types when constructing the dictionary. Table 4 shows instance names, answer nodes, and one or more constraint nodes in the query "*papers that were cited by papers that were written by John*".

Table 4. Graph nodes in the query "papers that were cited by papers that were written by John"

| Named Entity | Instance | Answer Node | Constraint Node |
|---|---|---|---|
| papers | cited_Class_Paper_1 | Yes | No |
| papers | citing_Class_Paper_2 | No | No |
| John | citing_Author_3 | No | Yes |

Information shown in Table 4 is an important building block of a graph query language used to query graph databases. It enables the construction of a graph query language by providing all necessary information of nodes in a bibliographic graph.

3.6.2. The check of connectedness and directions of graph relations

Connectedness denotes whether two bibliographic named entities are directly connected in a database schema. For example, two bibliographic named entities "*papers*" and "*happy university*" in the query "*papers by happy university*" are not directly connected in the schema: "*Paper*" is connected to "*Author*" and "*Author*" is connected to "*Organization*". Even though the parsing results suggest a dependency relation between the two bibliographic named entities, the dependency relation should not be selected as a graph relation because it does not conform to the database schema. Thus, we check every dependency relation and add required nodes and relations to form a complete set of graph relations (Figure 3).

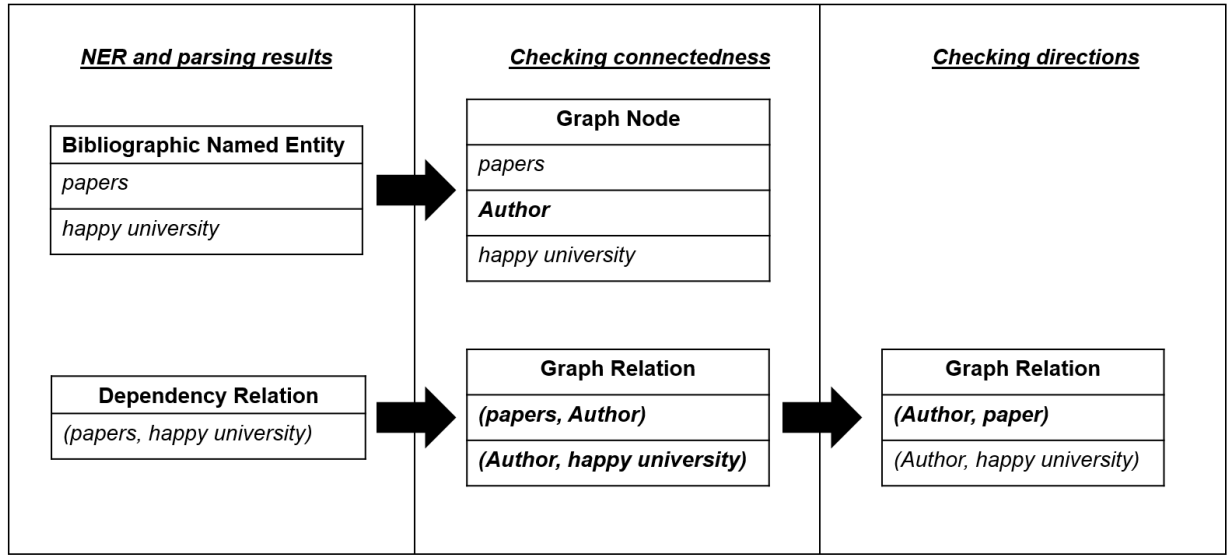

Figure 3: The check of connectedness and directions of the query "papers by happy university"

After checking the connectedness of each graph relation and adding necessary new nodes and relations, we check the direction of each graph relation to see whether the source and target of each graph relation conforms to the database schema. In a graph query language, we need to provide a set of graph relations with explicit definitions of sources and targets. For example, the relation between "*Paper*" and "*Author*" can be either modeled as "*WRITES*" or "*IS_WRITEEN_ BY*", which have different directions. In the above example, the graph relation *(papers, Author)* was converted into *(Author, papers)* based on the schema we used.

3.6.3. The integration of cited and citing parts

As mentioned previously, we divide a query that involves citations into two parts to reduce the complexities in interpreting natural language queries. These two parts are parsed and converted into graph nodes and graph relations separately. To generate a single graph query, we need to integrate both nodes and relations from two parts. The integration of nodes is achieved by creating a new node set and moving all cited and citing graph nodes to the set. The integration of relations is achieved by connecting two bibliographic named entities with the type of "*Paper*" in cited and citing parts. If one or two parts do not include a bibliographic named entity with the type of "*Paper*", we add a new graph node "*Paper*" to the part(s) and a graph relation that connects cited paper and citing paper. For example, the query "*authors cited by John*" denotes authors whose papers that were cited by papers written by John, but both the cited and citing part do not have a bibliographic named entity with the type of "*Paper*". Figure 4 shows the way to handle such queries.

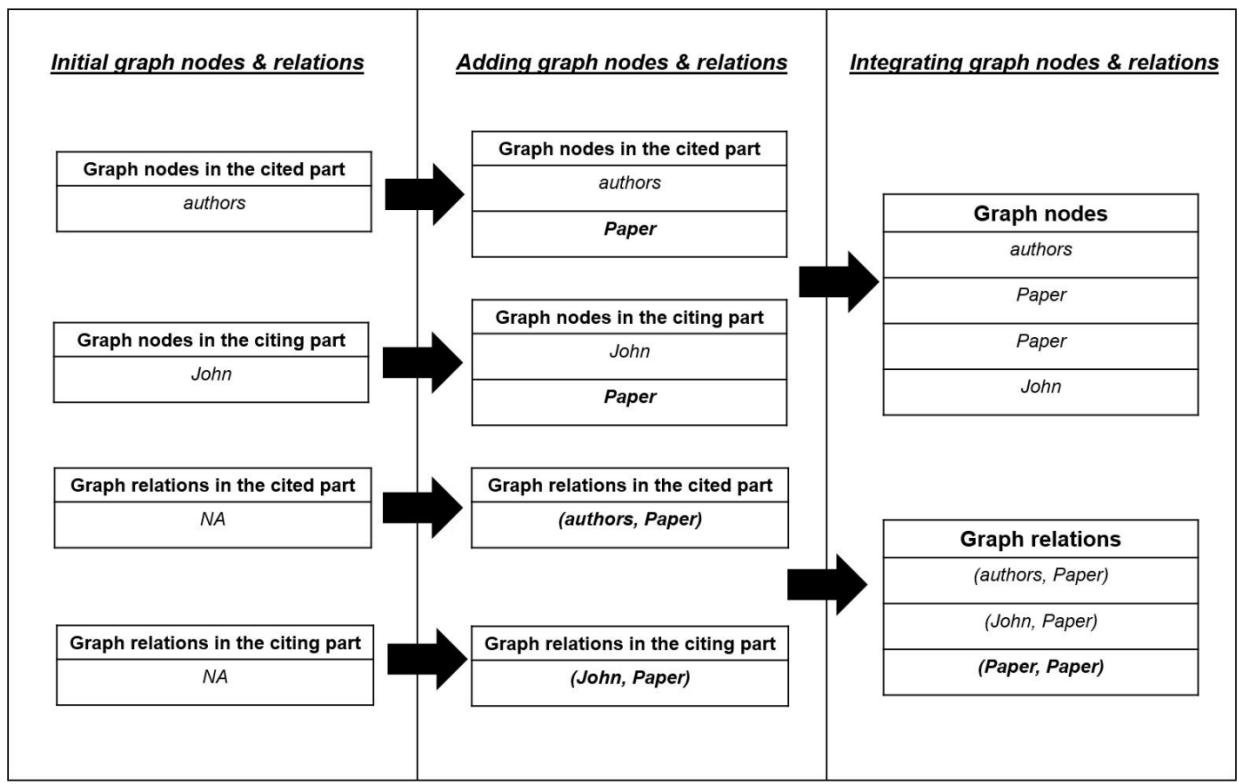

Figure 4. The integration of cited and citing parts in the query "authors cited by John"

As shown in Figure 4, two "*Paper*" nodes are added to both cited and citing parts. The nodes are then connected to the existing nodes "*authors'* and "*John*", respectively. Finally, two "*Paper*" nodes are connected through a citation relation.

*3.7. The translation of a graph query into a graph query language*

In this step, we translate a graph query into a graph query language. Widely used graph query languages such as Cypher, Germlin, and SPARQL have different syntaxes, but have the same building blocks, i.e., patterns, constraints, and return types. Because graph relations in a graph query are checked for connectedness and directions, and thus conform to the database schema, they can be directly translated into a graph query language. Constraints and return types are also available as we identify an answer node and constraint nodes in the previous step. Figure 5 shows how the graph query of a natural language query "*authors that were cited by John*" is translated into a graph query language. Four graph nodes derived from four named entities (NE1, NE2, NE3, and NE4) and three relations (R1, R2, and R3) among these graph nodes are identified. These nodes and relation are directly used to generate a graph query language.

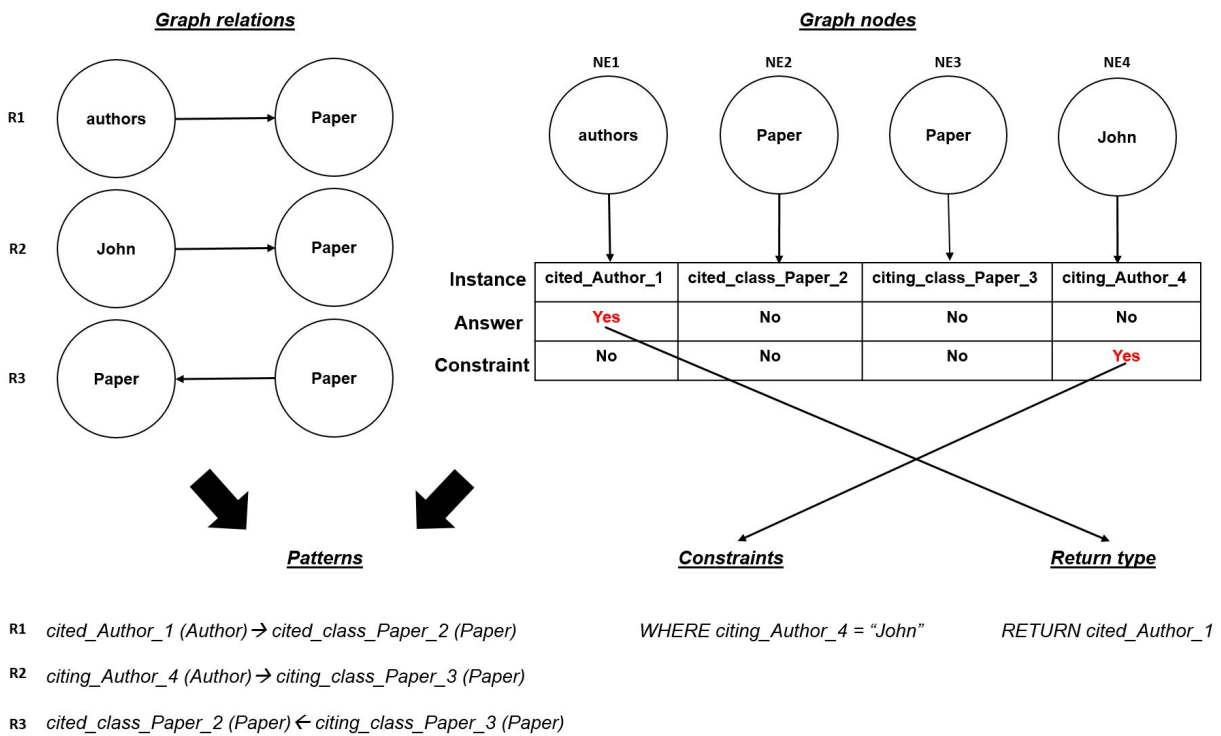

Figure 5. The translation of the graph query "authors that were cited by John" into a graph query language

Graph relations are used to derive patterns (i.e., paths), and a constraint is derived from the constraint node (i.e., *citing_Author_4*). The return type in a graph query language is the answer node (i.e., *cited_Author_1*) in the graph query. With these three building blocks, a query language can be generated.

*3. 8. The query of a graph database*

The generated query is submitted to a graph database to retrieve bibliographic data. Another option to query graph databases is to use embedded codes written in programming languages such as Java and C++, as graph databases provide application program interface (API) for data management. However, this approach would reduce the compatibility of a system because graph databases have different APIs. Thus, the framework is designed to translate a natural language query into a graph query language that is supported by a number of graph databases (Holzschuher & Peinl, 2013).

## 4. System Implementation and Experimental Results

*4.1. System implementation*

A web-based system is implemented by adding a natural language querying layer to a graph-based system (Zhu, Yan, & Song, 2016). It is based on the Spring Framework (as an application framework), Neo4j (as a graph database), and D3.js (as a visualization library). Figure 6 shows the graphical interface for users to formulate natural language queries. The example query is "*Papers about classification, which were cited by Asoke K. Nandi 's papers that had been presented in Pattern Recognition*".

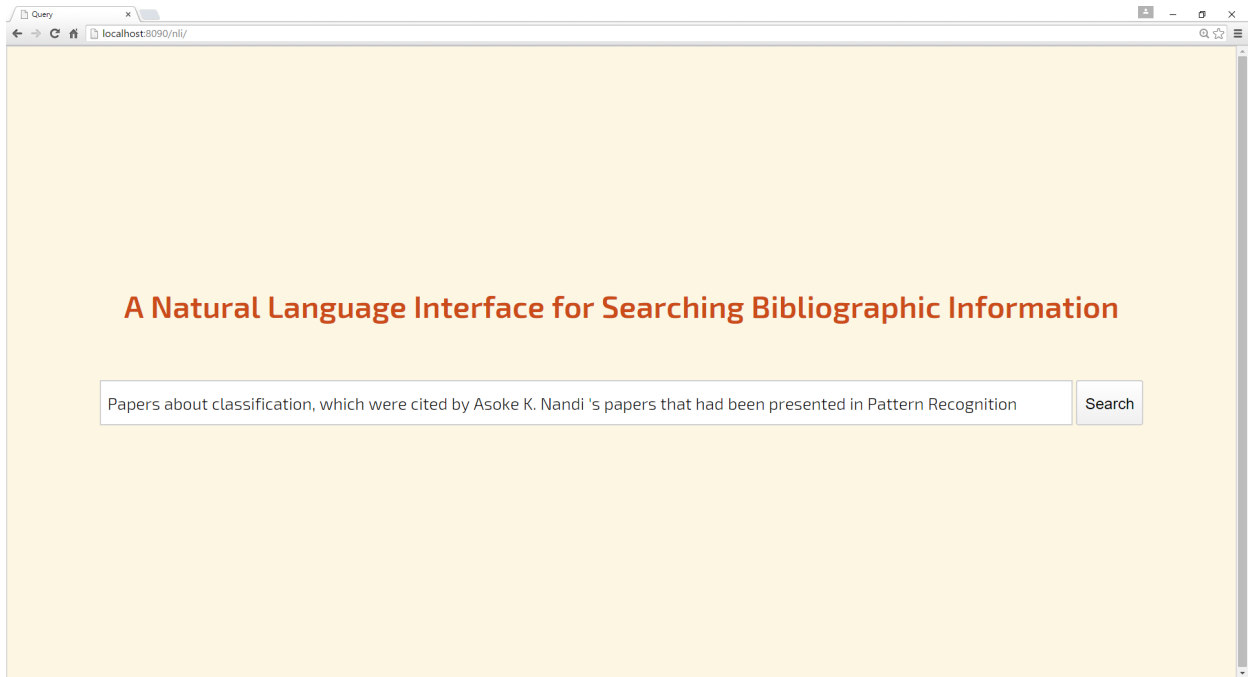

Figure 6. A natural language interface with an example query

After typing the natural language query and clicking the "*Search*" button, the system analyzes the natural language query. Recognized bibliographic named entities, dependency relations of the query, graph nodes, graph relations, and a graph query are shown in Figure 7.

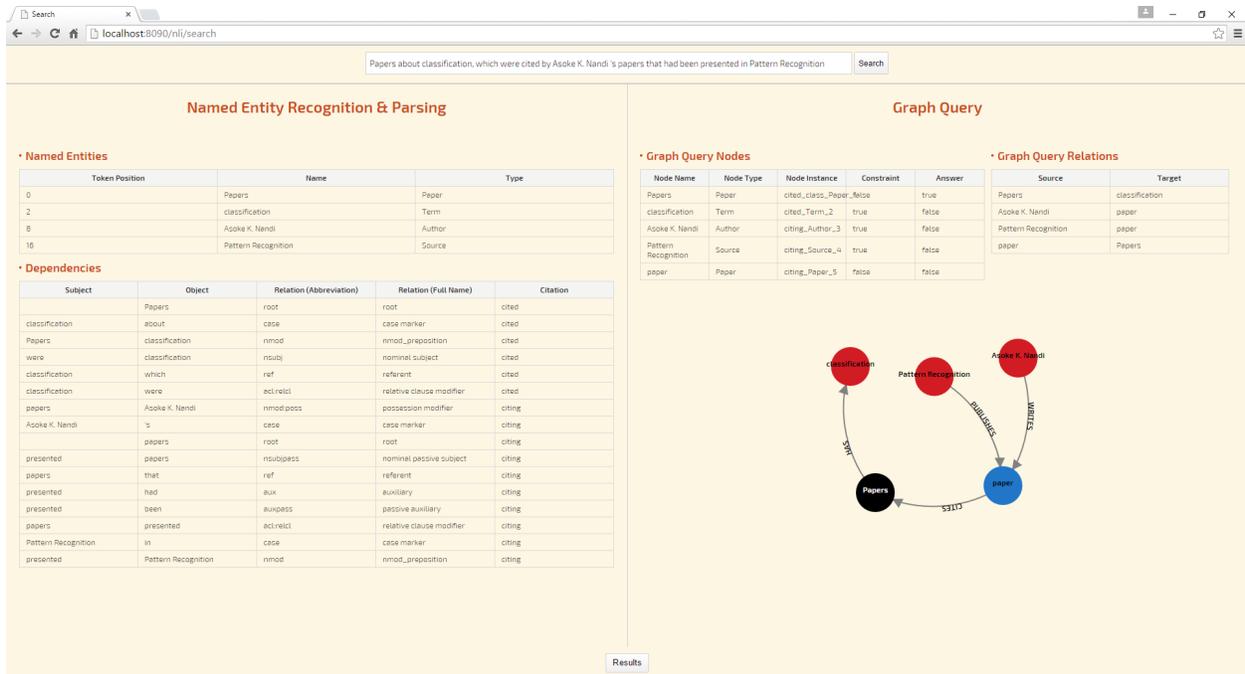

Figure 7. The analysis of a natural language query

Figure 7 shows how the example query was analyzed. First, bibliographic named entities such as *Papers*, *classification*, *Asoke K. Nandi, papers,* and *Pattern Recognition* were recognized. As mentioned previously, these bibliographic named entities were extracted from the dataset we used in the experiment and stored into a dictionary. Dependency relations among all tokens in the query are also shown as the result of a syntactic analysis. Nodes were then obtained from bibliographic named entities while relations were selected from dependency relations. By integrating graph nodes and graph relations, the system generated a graph query to visualize the results of the natural language query. As an interactive information retrieval system, users can modify or proceed with the current natural language query by referencing the analysis of the graph query. The final search results are obtained by clicking the "*Results*" button (Figure 8).

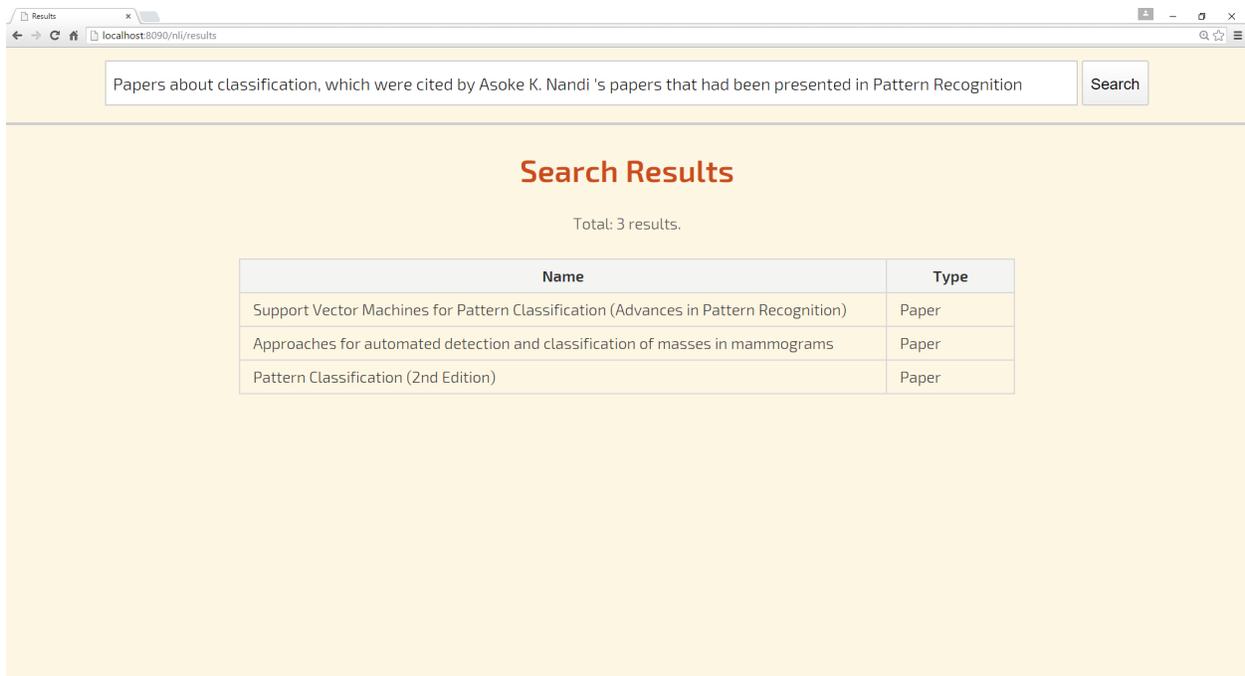

Figure 8. The search results of the example query

Figure 8 shows the final search results, which are the correct answers for the example query. In the system, we used Cypher as the graph query language, which is the default query language of Neo4j. Search results showed that there are three entries that matched the natural language query.

*4.2. Experimental Results*

The data used in the experiment were derived from a dataset provided by Tang and colleagues (2008). It contains 629,814 papers, 595,775 authors, 12,609 sources, 291,109 terms, and 1,000 organizations. Given the characteristics of a natural language interface, precision and recall are not suitable metrics because they are 100%, if the natural language query is interpreted correctly, and 0%, otherwise (Li & Jagadish, 2014). Thus, the effectiveness of a natural language interface to a relational database or a knowledge base is evaluated as the ratio of correctly answered queries and query execution time, as practiced in related research (e.g., Li & Jagadish, 2014; Tablan et al., 2008; Zhu, Yan, & Song, 2016).

We tested both the ratio of correctly answered queries and query execution time by forming four groups of queries based on the number of bibliographic named entities in a query, which ranges from two-named entities to five-named entities. Ten queries for each group were tested. When formulating test queries, we considered a variety of meta-paths and included as many meta-paths as possible. For example, for two-node queries, we included meta-paths such as "*Author→ Paper*", "*Author→ Organization*", "*Source→ Paper*", "*Paper→ Term*", and "*Paper→ Paper*". As the number of named entities in a query increases, the number of meta-paths also grows. Therefore, we selected 10 meta-paths that are representative in bibliographic searching based on our domain knowledge. Forty tested queries are listed in the Appendix. The ratio of correctly answered queries for each group is shown in Table 5.

Table 5. The ratio of correctly answered queries

| The number of named entities | 2 | 3 | 4 | 5 |
|---|---|---|---|---|
| The ratio of correctly answered queries | 10/10 | 10/10 | 9/10 | 10/10 |

As shown in Table 5, we did not see a correlation between the number of named entities in a query and the ratio of correctly answered queries. The example query that our framework processed incorrectly is "*Authors who are affiliated with University007 and wrote Papers about clustering*". The reason of the misinterpretation is that the parser misidentified "*wrote*" as the root of the query, which should be "*authors*". Our framework performed 100% correctly for all other test queries.

Query execution time includes the time of interpreting a natural language query (i.e., recognizing named entities and parsing) and the time of answering the query in a graph database. Time spent in formulating a query is not considered to leave out human factors and to focus on the performance of the system. The test environment is a laptop PC with a Windows 7 64-bit operating system, an Intel Core i5-3320M CPU, and 16GB RAM. The execution time for each query and average execution time in each group are shown in Figure 9. The query that was incorrectly interpreted was excluded from the calculation.

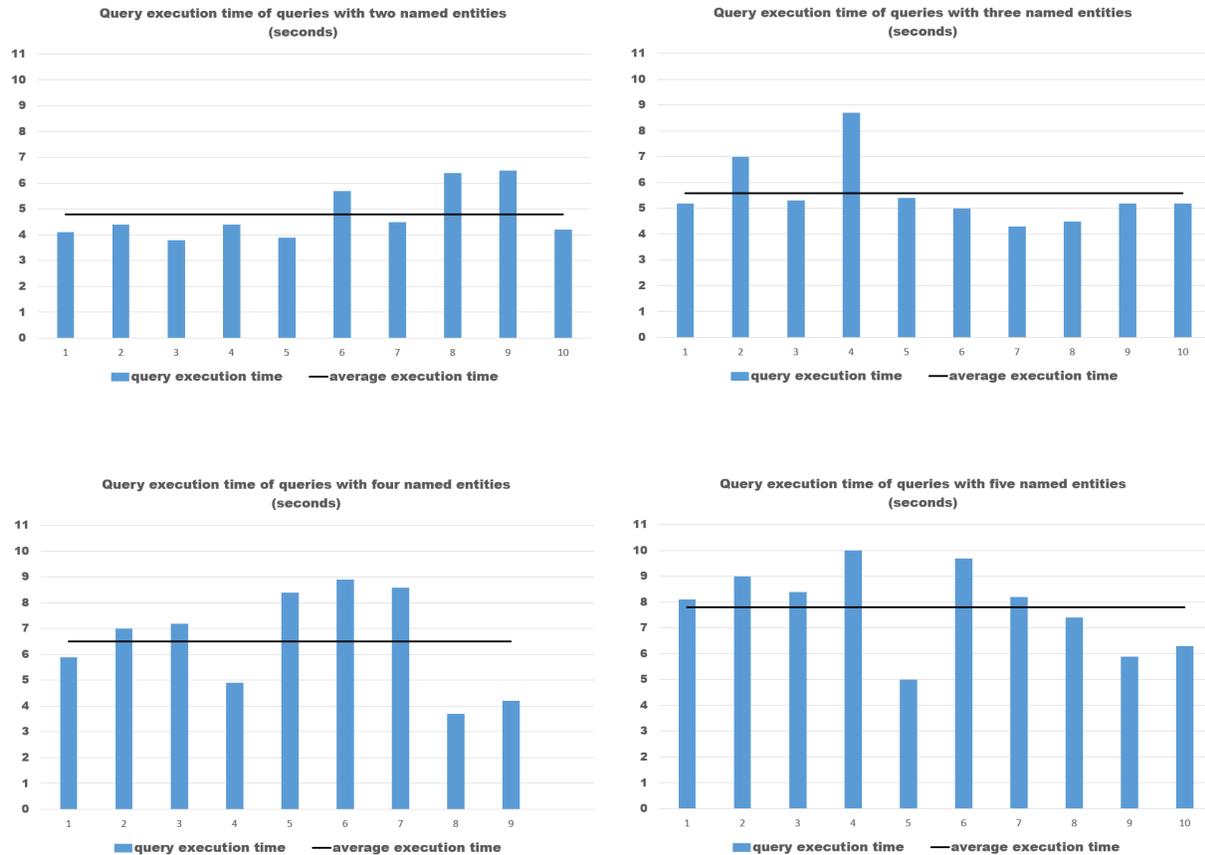

Figure 9. The query execution time of queries with the number of named entities from two to five

Query execution time is affected by the length of a query as well as the number of items in the search result that matched the query. A long natural language query need more time to be interpreted than a short query as the time spent on recognizing named entities in the query and parsing the query increases. Query execution time also increases if there are many items that matched the query. The average execution time is 4.8 seconds for two-named entity query, 5.6 seconds for three-named entity query, 6.5 seconds for four-named entity query, and 7.8 seconds for five-named entity query. The longest time taken to process a query is about ten seconds. Nonetheless, an industry-scale systems use more powerful servers, we believe the execution time should be reduced in real-world use cases.

## 5. Conclusions

In this paper, we presented a natural language interface for searching bibliographic data. We designed and implemented a framework of building a natural language interface to a graph-based bibliographic information retrieval system. The framework allows users to query bibliographic information by formulating and answering queries represented in natural language. An important step in interpreting natural language queries is to recognize bibliographic named entities in natural language queries. We identified relations among recognized bibliographic entities by parsing queries and finding dependency relations. We tested the framework using a large empirical dataset, and the experiment results showed that the method correctly interpreted 39 out of 40 natural language queries with varied levels of complexities.

The contributions of this paper are as follows: first, to our knowledge, it is the first attempt to propose a natural language interface to graph-based bibliographic information retrieval; second, we have proposed a novel customized natural language processing framework that integrates a few original algorithms/heuristics for interpreting and analyzing natural language bibliographic queries; and third, we have shown that the proposed framework and natural language interface provide a practical solution to build real-world natural language interface-based bibliographic information retrieval systems.

Our natural language interface has several limitations. First, it is domain-dependent. Because bibliographic information retrieval is a specialized area, some customized rules and heuristics were introduced in the framework to ensure higher performance, which might limit its applicability to other domains. In addition, since there has been no prior study in this area, there lacked a baseline, which did not allow us to compare the performance of our framework with previous studies. Instead, we created test queries with varying lengths and complexities from scratch by considering a variety of bibliographic information needs. Thus, from the perspective of evaluations, our contribution lies in the building of a baseline (i.e., a test dataset and benchmark scores) for future studies related to natural language-based bibliographic information retrieval. Another limitation of the study lies in the users' vocabulary use and entity resolution. Because the proposed framework focuses on a broad scope of a complete system design, it does not strongly tackle the above issues. Methods used in query expansion could be one of useful solutions for these issues, and we expect to address these issues in our future works. Lastly, we see the value of user centered evaluation and plan to evaluate the system with users in our future works.

Concepts and procedures introduced in this paper can serve as a foundation and guideline for future studies that aim to improve bibliographic information retrieval by utilizing the power of natural language. We expect natural language interfaces to retrieval systems would be significantly improved as technologies of natural language processing advances. With an ever-increasing volume of publications, a natural language interface is a promising solution to cope with this and guide users toward a more informed bibliographic information retrieval.

## Acknowledgement

This project was made possible in part by the Institute of Museum and Library Services (Grant Award Number: RE-07-15-0060-15), for the project titled "Building an entity-based research framework to enhance digital services on knowledge discovery and delivery".

## References

Abacha, A. B., & Zweigenbaum, P. (2015). MEANS: A medical question-answering system combining NLP techniques and semantic web technologies. Information Processing & Management, 51(5), 570-594. doi:10.1016/j.ipm.2015.04.006

Aghaei Chadegani, A., Salehi, H., Yunus, M. M., Farhadi, H., Fooladi, M., Farhadi, M., & Ale Ebrahim, N. (2013). A comparison between two main academic literature collections: Web of Science and Scopus databases. Asian Social Science, 9(5), 18-26.

Androutsopoulos, I., Ritchie, G. D., & Thanisch, P. (1995). Natural language interfaces to databases–an introduction. Natural language engineering, 1(01), 29-81.

Angles, R., & Gutierrez, C. (2008). Survey of graph database models. ACM Computing Surveys (CSUR), 40(1), 1.

Berners-Lee, T., Hendler, J., & Lassila, O. (2001). The semantic web. Scientific American, 284(5), 28-37.

Bikel, D. M., Miller, S., Schwartz, R., & Weischedel, R. (1997). Nymble: a high-performance learning name-finder. In Proceedings of the fifth conference on applied natural language processing (pp. 194-201). Association for Computational Linguistics.

Cafarella, M. J., & Etzioni, O. (2005, May). A search engine for natural language applications. In Proceedings of the 14th international conference on World Wide Web (pp. 442-452). ACM.

Chieu, H. L., & Ng, H. T. (2002). Named entity recognition: a maximum entropy approach using global information. In Proceedings of the 19th international conference on Computational linguistics-Volume 1 (pp. 1-7). Association for Computational Linguistics.

Cimiano, P., Haase, P., Heizmann, J., Mantel, M., & Studer, R. (2008). Towards portable natural language interfaces to knowledge bases – The case of the ORAKEL system. Data and Knowledge Engineering, 65(2), 325–354.

Codd, E. F. (1970). A relational model of data for large shared data banks. Communications of the ACM, 13(6), 377-387.

Collins, M. (1997). Three generative, lexicalised models for statistical parsing. In Proceedings of the 35th Annual Meeting of the Association for Computational Linguistics and Eighth Conference of the European Chapter of the Association for Computational Linguistics (pp. 16-23). Association for Computational Linguistics.

Damljanovic, D., Agatonovic, M., & Cunningham, H. (2010). Natural language interfaces to ontologies: Combining syntactic analysis and ontology-based lookup through the user interaction. In The semantic web: Research and applications (pp. 106-120). Springer Berlin Heidelberg.

Derczynski, L., Maynard, D., Rizzo, G., Erp, M. v., Gorrell, G., Troncy, R., Bontcheva, K. (2015). Analysis of named entity recognition and linking for tweets. Information Processing & Management, 51(2), 32-49. doi:10.1016/j.ipm.2014.10.006

De Marneffe, M. C., MacCartney, B., & Manning, C. D. (2006). Generating typed dependency parses from phrase structure parses. In Proceedings of LREC (Vol. 6, No. 2006, pp. 449-454).

Doszkocs, T. E., & Rapp, B. A. (1979). Searching MEDLINE in english: A prototype user interface with natural language query, ranked output, and relevance feedback. Information Choices and Policies: Proceedings of the 1979 ASIS Annual Meeting, Volume 16, New York, Knowledge Industry Publications for American Society for Information Science, 1979 131-139.s,

Earley, J. (1970). An efficient context-free parsing algorithm. Communications of the ACM, 13(2), 94-102.

Falagas, M. E., Pitsouni, E. I., Malietzis, G. A., & Pappas, G. (2008). Comparison of PubMed, Scopus, web of science, and Google scholar: strengths and weaknesses. The FASEB journal, 22(2), 338-342.

Fazzinga, B., & Lukasiewicz, T. (2010). Semantic search on the web. Semantic Web, 1(1–2), 89–96.

Fersini, E., Messina, E., Felici, G., & Roth, D. (2014). Soft-constrained inference for named entity recognition. Information Processing & Management, 50(5), 807-819. doi:10.1016/j.ipm.2014.04.005

Finkel, J. R., Grenager, T., & Manning, C. (2005). Incorporating non-local information into information extraction systems by gibbs sampling. In Proceedings of the 43rd Annual Meeting on Association for Computational Linguistics (pp. 363-370). Association for Computational Linguistics.

Gusfield, D. (1997). Algorithms on strings, trees, and sequences: Computer science and computational biology. New York; Cambridge [England];: Cambridge University Press.

Habernal, I., & Konopík, M. (2013). SWSNL: Semantic web search using natural language. Expert Systems with Applications, 40(9), 3649-3664.

Holzschuher, F., & Peinl, R. (2013). Performance of graph query languages: comparison of cypher, gremlin and native access in neo4j. In Proceedings of the Joint EDBT/ICDT 2013 Workshops (pp. 195-204). ACM.

Jacso, P. (2005). As we may search-Comparison of major features of the Web of Science, Scopus, and Google Scholar citation-based and citation-enhanced databases. CURRENT SCIENCE-BANGALORE-, 89(9), 1537.

Kaufmann, E., & Bernstein, A. (2010). Evaluating the usability of natural language query languages and interfaces to Semantic Web knowledge bases. Web Semantics: Science, Services and Agents on the World Wide Web, 8(4), 377-393.

Kaufmann, E., Bernstein, A., & Fischer, L. (2007). NLP-reduce: A ''naïve'' but domain independent natural language interface for querying ontologies. In 4th European Semantic Web Conference (ESWC 2007) (pp. 1–2).

Klein, D., & Manning, C. D. (2003). Accurate unlexicalized parsing. In Proceedings of the 41st Annual Meeting on Association for Computational Linguistics-Volume 1 (pp. 423-430). Association for Computational Linguistics.

Klein, D., & Manning, C. D. (2004). Corpus-based induction of syntactic structure: Models of dependency and constituency. In Proceedings of the 42nd Annual Meeting on Association for Computational Linguistics (p. 478). Association for Computational Linguistics.

Li, F., & Jagadish, H. V. (2014). Constructing an interactive natural language interface for relational databases. Proceedings of the VLDB Endowment, 8(1), 73-84.

Marcus, M. P., Marcinkiewicz, M. A., & Santorini, B. (1993). Building a large annotated corpus of English: The Penn Treebank. Computational linguistics, 19(2), 313-330.

McCallum, A., & Li, W. (2003). Early results for named entity recognition with conditional random fields, feature induction and web-enhanced lexicons. In Proceedings of the seventh conference on Natural language learning at HLT-NAACL 2003-Volume 4 (pp. 188-191). Association for Computational Linguistics.

Miller, E. (1998). An introduction to the resource description framework. Bulletin of the American Society for Information Science and Technology, 25(1), 15-19.

Mu, X., Lu, K., & Ryu, H. (2014). Explicitly integrating MeSH thesaurus help into health information retrieval systems: An empirical user study. Information Processing & Management, 50(1), 24-40. doi:10.1016/j.ipm.2013.03.005

Nadeau, D., & Sekine, S. (2007). A survey of named entity recognition and classification. Lingvisticae Investigationes, 30(1), 3-3. doi:10.1075/li.30.1.03nad

Noessner, J., Niepert, M., Meilicke, C., & Stuckenschmidt, H. (2010). Leveraging terminological structure for object reconciliation. In The Semantic Web: Research and Applications (pp. 334-348). Springer Berlin Heidelberg.

Park, C., & Lim, S. (2015). Efficient processing of keyword queries over graph databases for finding effective answers. Information Processing & Management, 51(1), 42.

Rau, L. F. (1991). Extracting company names from text. In Artificial Intelligence Applications, 1991. Proceedings. Seventh IEEE Conference on (Vol. 1, pp. 29-32). IEEE.

Rindflesch, T. C., Tanabe, L., Weinstein, J. N., & Hunter, L. (2000). EDGAR: extraction of drugs, genes and relations from the biomedical literature. In Pac Symp Biocomput (Vol. 5, pp. 514-25).

Roy, S., & Zeng, W. (2013). Cognitive canonicalization of natural language queries using semantic strata. ACM Transactions on Speech and Language Processing (TSLP), 10(4), 1-30. doi:10.1145/2539053

Ryu, P., Jang, M., & Kim, H. (2014). Open domain question answering using wikipedia-based knowledge model. Information Processing & Management, 50(5), 683-692. doi:10.1016/j.ipm.2014.04.007

Score, S. C. (2009). Web of Science and Scopus: A comparative review of content and searching capabilities. The Charleston Advisor.

Sun, Y., Yu, Y., & Han, J. (2009). Ranking-based clustering of heterogeneous information networks with star network schema. Paper presented at the 797-806. doi:10.1145/1557019.1557107

Tablan, V., Damljanovic, D., & Bontcheva, K. (2008). A natural language query interface to structured information. (pp. 361-375). Berlin, Heidelberg: Springer Berlin Heidelberg. doi:10.1007/978-3-540-68234-9_28

Tang, J., Zhang, J., Yao, L., Li, J., Zhang, L., & Su, Z. (2008). Arnetminer: extraction and mining of academic social networks. In Proceedings of the 14th ACM SIGKDD international conference on Knowledge discovery and data mining (pp. 990-998). ACM

Tsuruoka, Y., & Tsujii, J. I. (2003). Boosting precision and recall of dictionary-based protein name recognition. In Proceedings of the ACL 2003 workshop on Natural language processing in biomedicine-Volume 13 (pp. 41-48). Association for Computational Linguistics.

Tumer, D., Shah, M. A., & Bitirim, Y. (2009). An empirical evaluation on semantic search performance of keyword-based and semantic search engines: Google, Yahoo, MSN and Hakia. In Internet Monitoring and Protection, 2009. ICIMP'09. Fourth International Conference on (pp. 51-55). IEEE.

Vicknair, C., Macias, M., Zhao, Z., Nan, X., Chen, Y., & Wilkins, D. (2010). A comparison of a graph database and a relational database: a data provenance perspective. In Proceedings of the 48th annual Southeast regional conference (p. 42). ACM.

Wang, C., Xiong, M., Zhou, Q., & Yu, Y. (2007). Panto: A portable natural language interface to ontologies. In The Semantic Web: Research and Applications (pp. 473-487). Springer Berlin Heidelberg.

Zhu, Y., Yan, E., & Song, I.-Y. (2016). The use of a graph-based system to improve bibliographic information retrieval: System design, implementation, and evaluation. Journal of the Association for Information Science and Technology. doi: 10.1002/asi.23677

**Appendix: Natural language queries tested in the experiment**

1. Papers by Gerard Salton
2. Michael Lawrence's papers
3. Papers that were written by Sangjun Lee
4. Papers about ontology
5. Authors of Automatic text structuring experiments
6. Papers that were cited by Energy-Aware and Time-Critical Geo-Routing in Wireless Sensor Networks
7. Terms of Opacity generalised to transition systems
8. Organization of Johann Eder
9. Sources that published The Effect of Faults on Network Expansions
10. Papers that were published in Theoretical Computer Science
11. Papers about classification and DNA
12. Papers that were written by John R. Mick and published in ACM SIGMICRO Newsletter
13. Papers cites papers that were written by Braham Barkat

14. Papers about modulation which were published in Neural Networks
15. Authors of University713 who wrote A control word model for detecting conflicts between microoperations
16. Sources that published Zesheng Chen's papers
17. Authors whose papers were published in AI Communications
18. Authors who wrote papers that were about simulation
19. Terms of Junghyun Nam's papers
20. Organizations of authors of A New Quadtree Decomposition Reconstruction Methods
21. Papers about survey, semantic, and retrieval
22. Authors of papers that were cited by papers that were published in Decision Support Systems
23. Papers that cite papers that were written by Rainer Engelke and published in Microsystem Technologies
24. Nina Yevtushenko's papers that were cited by papers that were written by Sergey Buffalov
25. Sources that published papers about genome and mining
26. Terms of Rafae Bhatti's papers that were published in Communications of the ACM
27. Sources that published Tomasz Jurdzinski's papers which are about automata
28. Terms of papers that were written by authors at University123
29. Organizations of authors whose papers were published in Journal of Multivariate Analysis
30. Authors who are affiliated with University007 and wrote papers about clustering
31. Papers about classification, which were cited by Asoke K. Nandi 's papers that had been presented in Pattern Recognition
32. Authors of papers that were cited by papers that were written by Changqiu Jin and published in Journal of Computational Physics
33. Terms of papers that were cited by papers about kernel and regression
34. Sources that published papers cited papers about middleware and embedded

35. Organizations of authors whose papers were cited by papers that were published in Journal of Robotic Systems
36. Organizations of authors who wrote paperson similarity and bayesian
37. Papers about bayesian and electron which were written by authors at University170
38. Sources of papers, which were about eigenvalue and written by authors at University40
39. Authors at University899, who wrote papers that were about classifier, which were published in Applied Intelligence
40. Terms of papers that were published in Cybernetics and Systems Analysis and written by authors at University362